\documentclass[aps,prb,twocolumn,showpacs,superscriptaddress]{revtex4}
\usepackage{graphicx}
\usepackage{hyperref}
\usepackage{natbib}
\begin{document}
\title{Effect of local Coulomb interactions on the electronic structure and
exchange interactions in Mn$_{12}$ magnetic molecules}
\author{D. W. Boukhvalov}
\affiliation{Forschungszentrum Juelich, D-52425 Julich, Germany}
\affiliation{Institute of Metal Physics, Ekaterinburg 620219, Russia}
\author{A. I. Lichtenstein}
\affiliation{University of Nijmegen, NL-6525 ED Nijmegen, the Netherlands}
\author{V. V. Dobrovitski}
\affiliation{Ames Laboratory, Iowa State University, Ames IA 50011, USA}
\author{M. I. Katsnelson}
\affiliation{Institute of Metal Physics, Ekaterinburg 620219, Russia}
\affiliation{University of Nijmegen, NL-6525 ED Nijmegen, the Netherlands}
\affiliation{Ames Laboratory, Iowa State University, Ames IA 50011, USA}
\author{B. N. Harmon}
\affiliation{Ames Laboratory, Iowa State University, Ames IA 50011, USA}
\author{V. V. Mazurenko}
\affiliation{Institute of Metal Physics, Ekaterinburg 620219, Russia}
\affiliation{University of Nijmegen, NL-6525 ED Nijmegen, the Netherlands}
\author{V. I. Anisimov}
\affiliation{Institute of Metal Physics, Ekaterinburg 620219, Russia}

\begin{abstract}
We have studied the effect of local Coulomb interactions on the electronic
structure of the molecular magnet Mn$_{12}$-acetate within the LDA+U
approach. The account of the on-site repulsion results in a finite
energy gap and an integer value of the molecule's magnetic moment, both
quantities being in a good agreement with the experimental results. The
resulting magnetic moments and charge states of non-equivalent manganese
ions agree very well with experiments. The calculated values of the
intramolecular exchange parameters depend on the molecule's spin
configuration, differing by 25--30\% between the ferrimagnetic ground state
and the completely ferromagnetic configurations. The values of the
ground-state exchange coupling parameters are in reasonable agreement with
the recent data on the magnetization jumps in megagauss magnetic fields.
Simple estimates show that the obtained exchange parameters can be applied,
at least qualitatively, to the description of the spin excitations in Mn$%
_{12}$-acetate.
\end{abstract}

\maketitle

\section*{Introduction}

Single-molecule magnets represent a new class of ``zero-dimensional''
magnetic materials consisting of almost non-interacting, identical
point-like nanoscale entities \cite
{Molmag,Molmagn1,Molmagn2}. These magnetic molecules have
attracted a great deal of
attention from physicists as well as chemists for detailed
studies of many novel aspects of mesoscale magnetism, such as
quantum spin tunneling, spin relaxation of mesoscopic magnets, the
implications of the 
topological spin phase (Berry phase), etc. 
\cite{QTM,QTM1,QTM2,QTM3,QTM4,QTM5,stamp}. The 
Mn$_{12}$O$_{12}$[CH$_{3}$COO]$_{16}\cdot$4H$_{2}$O 
compound \cite{lis,sesnat}, often referred to as 
Mn$_{12}$-acetate or simply Mn$_{12}$, is one of the best-studied 
examples of this
kind of materials. Relatively slow dynamical processes 
taking place in not too
strong magnetic fields, when the intramolecular spin excitations can be
ignored, have been widely studied in experiments, and the properties of the
ground-state multiplet $S=10$ are known very well. Correspondingly, most of
the theoretical treatments of Mn$_{12}$ have been based on the rigid-spin
model (see e.g.\ Refs.\ 
\onlinecite{rigid,rigid1,rigid2,rigid3,rigid4,rigid5,rigid6,%
rigid7,rigid8,rigid9}),
assuming that the molecule's magnetic subsystem can be represented as a
single collective spin $S=10$. On the other hand, the internal magnetic and
electronic structure of the Mn$_{12}$ molecules and the intramolecular
interactions responsible for the formation of the molecule's collective spin
have been studied in less detail. 
Several experiments including inelastic
neutron scattering \cite{neutron}, megagauss field magnetization
measurements \cite{zvezdin}, and optical absorption \cite{optics} have been
performed to clarify the internal properties of the Mn$_{12}$ molecules. The
experimental data have been analyzed within the spin-Hamiltonian approach 
\cite{Mn12,zvezdin}, however, the resulting information was not systematic,
and often did not lead to unequivocal conclusions (see e.g.\
the discussion
of different parameter sets in Ref.\ \onlinecite{Mn12}).

Another approach to investigate intramolecular properties is {\it ab
initio} electronic structure calculations. So far, density-functional
computations of the electronic structure for Mn$_{12}$ have been carried out
within the local density approximation (LDA) \cite{ellis} and generalized
gradient approximation (GGA) \cite{Ped1,Ped2}. It is 
well known that for
transition metal-oxide systems the standard density functional methods (LDA
or GGA) do not give reliable information on the energy gaps and magnetic
superexchange interactions \cite{oxide,oxide1,oxide2,oxide3,oxide4}, and
many-body effects should be taken into account at least in the simplest form
of the 
LDA+U approximation \cite{LDA+U}. This is essential, e.g., for an
adequate description of\ the electronic structure and the spin excitation
spectra in MnO \cite{MnO,MnO1}. Mn$_{12}$-acetate is also an insulating
metal-oxide system with an energy gap of order \cite{optics} of 2 eV,
whereas the existing calculations \cite{ellis,Ped1,Ped2} give a finite
density of states (DOS) and small pseudogap at the Fermi level. As a result,
the total spin per molecule turns out to be non-integer, in contradiction
with experimental results. Futhermore, the intramolecular exchange couplings
in Mn$_{12}$ have never been calculated before. The GGA calculations have
recently been performed \cite{v15exch} for another single-molecule 
magnet, V$_{15}$,
but the resulting values had to be reduced by a factor of three
to achieve an agreement with the experiments. This difference could also be
caused by the neglect of the on-site Coulomb repulsion, since the latter is
known to decrease the calculated
exchange interactions in the metal-oxide compounds 
\cite{MnO,MnO1}.

In this work, we present the results of first-principle LDA+U calculations
of the electronic structure and exchange interactions for Mn$_{12}$. We show
that the account of the many-body effects within the LDA+U approximation
gives a reasonable
picture of the electronic structure, with a finite energy gap
and an 
integer value of the total spin of the molecule. Moreover, for the first
time we present the calculated data for the intramolecular exchange
constants in Mn$%
_{12}$. We discuss the results of our calculations and compare them with the
existing experimental data and previous theoretical studies.

The rest of the paper is organized as follows. In Section I we describe the
computational scheme and the approximations employed in this work. In
Section II the resulting electronic structure and calculated
magnetic moments in Mn$%
_{12}$ are presented. Section III is devoted to the computation of
the intramolecular exchange interactions. The conclusions are presented in
Section IV.

\section{Scheme of calculations}

\begin{figure}[b]
\includegraphics[width=7.5cm]{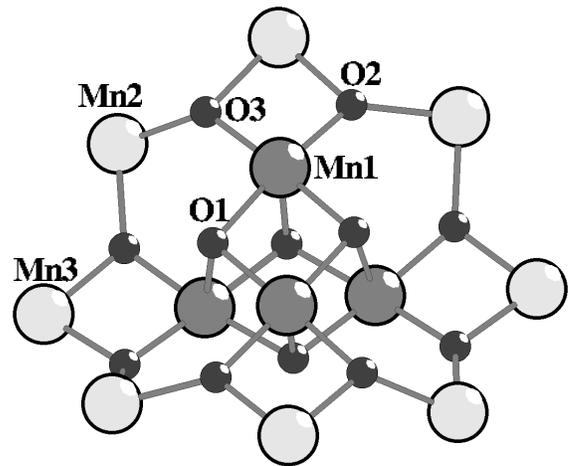}
\caption{The sketch of the arrangement of the manganese and 
oxygen ions in
the Mn$_{12}$ molecule. Big dark-grey circles represent 
the Mn$^{4+}$ ions,
big light-grey circles --- Mn$^{3+}$ ions, and small 
dark circles are
the bridging oxygens. Non-equivalent manganese ions 
Mn1, Mn2, Mn3,
and the oxygens O1, O2, O3 bridging them are indicated.}
\label{figtot}
\end{figure}

The calculations have been performed for the crystal structure found by
T.~Lis 
\cite{lis} with the simplifications similar to those made in Ref.\
\onlinecite{Ped2}
(all the methyl groups CH$_{3}$ have been replaced by hydrogen atoms and the
water molecules were omitted). Therefore, the real calculations have been
done for the system Mn$_{12}$O$_{12}$(HCOO)$_{16}$. Part of this
system is sketched in Fig.\ \ref{figtot}, showing the 
arrangement of
the manganese ions and the oxygens bridging them.

To start the LDA+U calculations one needs to identify the regions where the
atomic characteristics of the electronic states have largely survived
(``atomic spheres''). Within these atomic spheres one can expand an electron
wave function in terms of a localized orthonormal basis 
$\mid inlm\sigma \rangle $ (%
$i$ denotes the site, $n$ is the main quantum number, $l$ --- orbital quantum
number, $m$ --- magnetic number and $\sigma $ --- spin index). 
The density matrix is defined as 
\begin{equation}
n_{mm^{\prime }}^{\sigma }=-\frac{1}{\pi }\int^{E_{F}} {\text{Im}}\, 
 G_{inlm,inlm^{\prime }}^{\sigma }(E)dE,  \label{Occ}
\end{equation}
where $G_{inlm,inlm^{^{\prime }}}^{\sigma }(E)=\langle inlm\sigma \mid (E-%
\widehat{H})^{-1}\mid inlm^{^{\prime }}\sigma \rangle $ are the elements of
the Green function matrix in this localized representation 
(the effective Hamiltonian $\widehat{H%
}$ will be defined later). In terms of the elements of this density matrix 
$\{n^{\sigma }\}$, the generalized LDA+U functional\cite{LDA+U} has the
following form
\begin{eqnarray}
\label{U1}
E^{\text{LDA+U}}[\rho ^{\sigma }({\bf r}),\{n^{\sigma }\}] &=& 
E^{LSDA}[\rho ^{\sigma }({\bf r)}]\\
 \nonumber
&+&E^{U}[\{n^{\sigma }\}]- E_{dc}[\{n^{\sigma }\}]
\end{eqnarray}
Here $\rho ^{\sigma }({\bf r})$ is the charge density for 
the electrons with the spin $\sigma$,
and $E^{LSDA}[\rho ^{\sigma }({\bf r})]$ is the standard LSDA
functional. Eq. (\ref{U1}) asserts that the LSDA suffices in the absence of
orbital polarizations, which are given by
\begin{eqnarray}
\nonumber
E^{U}[\{n\}]&=&\frac{1}{2}\sum_{\{m\},\sigma }\langle m,m^{\prime \prime
}\mid V_{ee}\mid m^{\prime },m^{\prime \prime \prime }\rangle n_{mm^{\prime
}}^{\sigma }n_{m^{\prime \prime }m^{\prime \prime \prime }}^{-\sigma }\\
&+&  
(\langle m,m^{\prime \prime }\mid V_{ee}\mid m^{\prime },m^{\prime \prime
\prime }\rangle \\
 \nonumber
&-& \langle m,m^{\prime \prime }\mid V_{ee}\mid m^{\prime
\prime \prime },m^{\prime }\rangle )n_{mm^{\prime }}^{\sigma }n_{m^{\prime
\prime }m^{\prime \prime \prime }}^{\sigma },
\label{upart}
\end{eqnarray}
where $V_{ee}$ is the screened electron-electron interactions. The matrix
elements of $V_{ee}$ are defined via an average Coulomb parameter $U$
and the Hund intra-atomic exchange constant \cite{LDA+U} $J$. Finally, the
last term in Eq.\ (\ref{U1}) describes the correction for the double counting
(in the absence of orbital polarizations, Eq. (\ref{U1}) should reduce to $%
E^{LSDA}$), and is given by \cite{LDA+U}
\begin{eqnarray}
E_{dc}[\{n^{\sigma }\}]&=&\frac{1}{2}UN(N-1)-\frac{1}{2}J[N^{\uparrow
}(N^{\uparrow }-1)\\
 \nonumber
 &+& N^{\downarrow }(N^{\downarrow }-1)],
\end{eqnarray}
were $N^{\sigma }=\text{Tr}\,(n_{mm^{\prime }}^{\sigma })$ and $N=N^{\uparrow
}+N^{\downarrow }$.

In addition to the usual LDA potential, we find an effective single-particle
potential 
\begin{eqnarray}
\label{Pot}
V_{mm^{\prime }}^{\sigma }&=&\sum_{m^{\prime },m^{\prime \prime }}\{\langle
m,m^{\prime \prime }\mid V_{ee}\mid m^{\prime },m^{\prime \prime \prime
}\rangle n_{m^{\prime \prime }m^{\prime \prime \prime }}^{-\sigma } \\ 
 \nonumber
&+&(\langle m,m^{\prime \prime }\mid V_{ee}\mid m^{\prime },m^{\prime \prime
\prime }\rangle \\
 \nonumber
&-&\langle m,m^{\prime \prime }\mid V_{ee}\mid m^{\prime
\prime \prime },m^{\prime }\rangle )n_{m^{\prime \prime }m^{\prime \prime
\prime }}^{\sigma }\}- \\ 
 \nonumber
&-&U(N-\frac{1}{2})+J(N^{\sigma }-\frac{1}{2})
\end{eqnarray}
to be used in the effective single-particle Hamiltonian
\begin{equation}
\label{hamilt}
\widehat{H}=\widehat{H}_{LSDA}+\sum_{mm^{\prime }}\mid inlm\sigma \rangle
V_{mm^{\prime }}^{\sigma }\langle inlm^{\prime }\sigma \mid.
\end{equation}
The Linearized Muffin-Tin Orbitals (LMTO) method in the orthogonal
approximation\cite{lmto1} has been used for the LSDA calculations.

Using our Green's function method we can obtain the effective exchange
interaction parameters via the energy variation at small spin rotations 
\cite{KL}. The corresponding expression in the LDA+U scheme takes the 
form \cite{LAZ} 
\begin{equation}
\label{exchange}
J_{ij}=\sum_{\{m\}}I_{mm^{\prime }}^{i}\chi _{mm^{\prime }m^{\prime \prime
}m^{\prime \prime \prime }}^{ij}I_{m^{\prime \prime }m^{\prime\prime
\prime}}^{j},
\end{equation}
where the spin-dependent potentials $I$ are expressed in terms of the
potentials of Eq.\ (\ref{Pot}), 
\begin{equation}
I_{mm^{\prime }}^{i}=V_{mm^{\prime }}^{i\uparrow }-V_{mm^{\prime
}}^{i\downarrow },  \label{magpot}
\end{equation}
while the effective intersublattice susceptibilities ($i$ and $j$)are
defined in terms of the LDA+U eigenfunctions $\psi$ as 
\begin{equation}
\chi _{mm^{\prime }m^{\prime \prime }m^{\prime \prime \prime }}^{ij}=\sum_{%
{\bf knn}^{\prime }}\frac{n_{n{\bf k\uparrow }}-n_{n^{\prime }{\bf %
k\downarrow }}}{\epsilon _{n{\bf k\uparrow }}-\epsilon _{n^{\prime }{\bf %
k\downarrow }}}\psi _{n{\bf k\uparrow }}^{ilm^{\ast }}\psi _{n{\bf k\uparrow 
}}^{jlm^{\prime \prime }}\psi _{n^{\prime }{\bf k\downarrow }}^{ilm^{\prime
}}\psi _{n^{\prime }{\bf k\downarrow }}^{jlm^{\prime \prime \prime \ast }}.
\label{suscep}
\end{equation}
As a ground state, the standard spin arrangement has been chosen with the
spins of Mn1 ions directed down and the spins of Mn2 and Mn3 ions directed
up. This arrangement was self-consistent. To check the applicability of
the localized spin models to Mn$_{12}$ the calculations for the 
ferromagnetic configurations (all spins up) have also been performed.

\section{Electronic structure and magnetic moments}

Detailed calculations have been carried out for the three values of $U=4,6$,
and $8$ eV keeping constant the value of the intra-atomic
Hund's exchange parameter $J=0.9$
eV, to check the sensitivity of the results with respect to variation in the
local Coulomb repulsion $U$. For all values of $U$ used in the calculations,
the system turns out insulating with the energy gap between
$1.5-2$ eV, and
the integer moment of the ferrimagnetic ground state $\mu _{\text{total}}=20$
$\mu _{B}$. Dependence of the calculated quantities on the value of $U$ can
be inferred from the data in the Table \ref{momreal}, and as one can see,
for most of the properties the dependence on $U$ is reasonably weak. The
value $U=8$ eV seems to be slightly preferrable; it is known that this value
gives good agreement with experiments for other manganese-oxide systems, and
agrees reasonably 
well with the independent theoretical calculation \cite{MnO}
giving the value of 
$U=6.9$ eV. Therefore, below we discuss in detail only the case
of $U=8$ eV.

\begin{figure}[b]
\includegraphics[width=7.5cm]{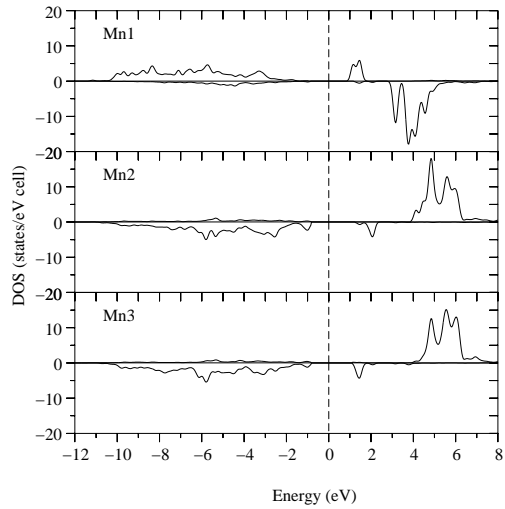}
\includegraphics[width=7.5cm]{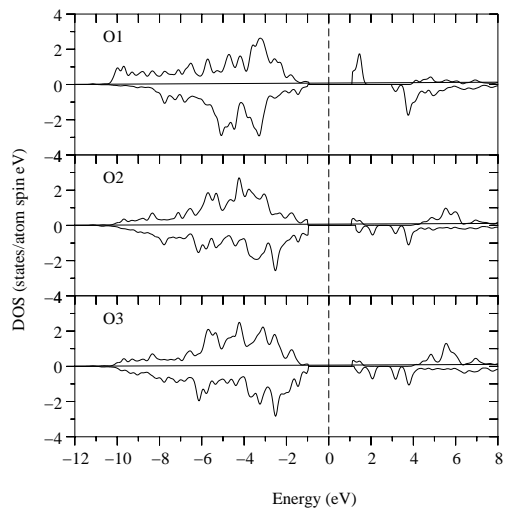}
\caption{Partial densities of states for different 
manganese ($d$ orbitals)
and oxygen ($p$ orbitals) atoms within the LDA+U approach. 
The values $U=8$
eV and $J=0.9$ eV have been used.}
\label{figldau}
\end{figure}

\begin{figure}[b]
\includegraphics[width=7.5cm]{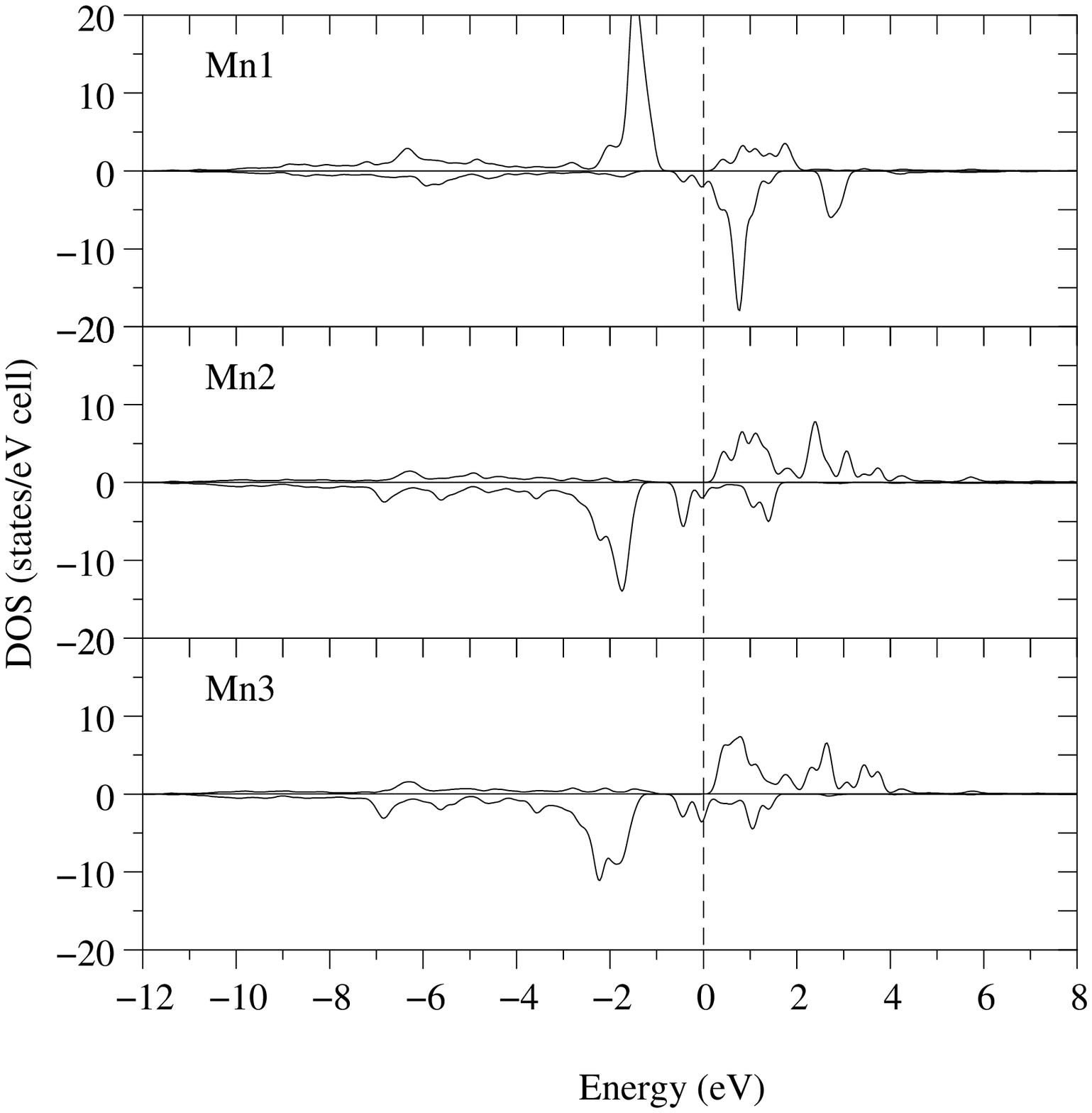}
\includegraphics[width=7.5cm]{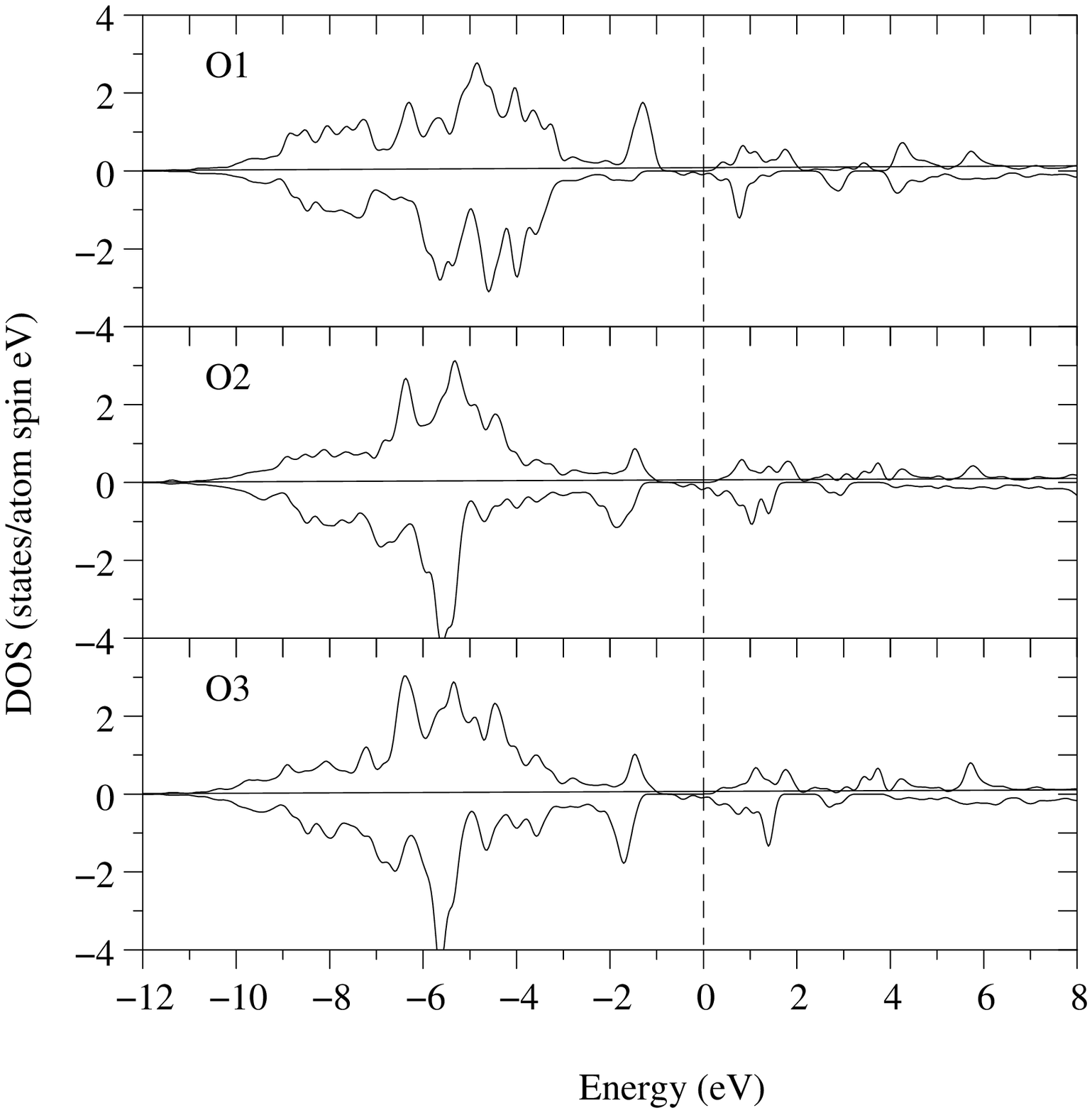}
\caption{Partial densities of states for different manganese ($d$
orbitals) and oxygen ($p$ orbitals) atoms calculated within the
LDA approach.}
\label{figlda}
\end{figure}

To begin with, let us compare the results given by LDA and LDA+U approaches.
In Figs.\ \ref{figldau} and \ref{figlda}, we show the 
calculated partial densities of states (DOS) for different manganese 
atoms and for the oxygen atoms belonging to the exchange
bridges. Our LDA
curves for 
the DOS are similar to those obtained in Refs.\ \onlinecite{Ped1,Ped2}
using 
the GGA approach. The main distinction of the LDA+U results as compared
with the LDA is the nonzero energy gap instead of the finite DOS at the
Fermi level. Moreover, there is a difference in the DOS peaks structure. All
manganese ions in Mn$_{12}$ molecule are in distorted octahedral
coordination, correspondingly, one can see the three peaks of $t_{2g}$
character situated below the two $e_g$-like peaks. For 
the LDA calculations the
spectral density is strongly concentrated within the region $\pm 2$ eV near
the Fermi level, see Fig.\ \ref{figlda}, while for 
the LDA+U calculations, the DOS
is spread rather uniformly over the region of about 10 eV.

It is known that the
LDA+U approach is effective for consideration of charge
separation and charge ordering phenomena \cite{LDA+U}. Mn$_{12}$ molecules
also exhibit Mn$^{3+}$ -- Mn$^{4+}$ disproportionality, 
which has been
directly demonstrated by the measurements of the X-ray absorption spectra 
\cite{mnxps}. The majority spin DOS curves (see Fig.\ \ref{figldau})
demonstrate three occupied $t_{2g}$ peaks and two empty $e_g$ peaks for Mn1,
while for Mn2 and Mn3 ions only one $e_g$ peak is empty; this picture indeed
corresponds to the charge disproportionality detected in the experiments.

Several remarks are in order. There are indications that the
manganese-oxygen bonds are considerably covalent, and $3d$ states of
manganese are highly hybridized with $2p$ states of oxygen. As can be seen
in Fig.\ \ref{figldau}, the main features in the partial DOS of manganese
ions have counterparts in the partial DOS of oxygen ions. Another indication
of the strong covalence is the discrepancy between the sum of the local $3d$%
-moments of Mn ions and the total magnetic moment of the molecule which is
calculated from the total DOS. The difference, which is of order of $2\mu_B$%
, can be attributed to the partial moments of oxygens estimated to be of
order of $0.05 \mu_B$.

The value of the energy gap calculated within LDA+U scheme varies from 1.78
eV to 2.01 eV with the change in the value of $U$ from 6 eV to 8 eV. Recent
optical measurements \cite{optics} show the features in the absorption and
reflection spectra in the region of the charge-transfer excitations at 1.7
eV, 2.0 eV, and 2.4 eV. Magneto-optical measurements \cite{cheesman} also
show the latter feature. These values are in very good agreement with the
calculated energy gap.

\section{Intramolecular exchange interactions}

At present, there is no reliable quantitative information about the
intramolecular exchange couplings in the Mn$_{12}$ molecules, making
difficult the validation of the corresponding computational results. Crude
estimates for the exchange couplings, based on the comparison with different
manganese-oxygen compounds, have been presented in Ref.\ 
\onlinecite{sessoli}. Quantitative evaluation of the exchange 
constants has been
performed in Ref. \onlinecite{zvezdin} using the results of the 
megagauss-field
experiments, however the magnetization curves appeared to be not very
sensitive to the values of the exchange parameters. Another quantitative
approach, based on the inelastic neutron scattering data, has been
undertaken \cite{Mn12}, but a simplified eight-spin model was used
which
makes the fitting not conclusive (see the discussion of various
possible sets of parameters in Ref. \onlinecite{Mn12}). 
Therefore, we believe that
it makes sense to focus the discussion of our results on qualitative trends
in the calculated parameters instead of their exact values.

\begin{figure}[b]
\includegraphics[width=7.5cm]{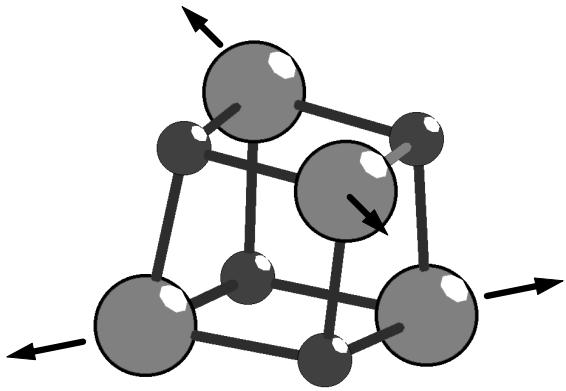}
\caption{The sketch of the arrangement of the manganese and 
oxygen ions in
the inner cubane of Mn$_{12}$ molecule. Arrows show the 
directions
of the displacement of the Mn ions which lead to the 
formation of the
inner cubane instead of the perfect cube.}
\label{figcub}
\end{figure}

The exchange parameters have been calculated for different values of $U$,
see Table \ref{momreal}. As expected for the superexchange mechanism, the
values of the couplings decrease with increasing $U$; again, below we
discuss only the values corresponding to $U=8$ eV. We have compared the
calculated values of the exchange parameters with the data suggested in
Refs.\ \onlinecite{sessoli,zvezdin}, see the Table \ref{exchreal}.
In this Table,
we have shown only the interactions between relatively closely situated
manganese ions; the exchange parameters between distant ions do not exceed
2--3 K. As one can see, the values obtained in our calculations are 
closer to
those resulting from the analysis of the megagauss-field experiments, while
the data suggested from general chemical consideration are bigger by about a
factor of three. However, the more interesting fact is that the exchanges
between the Mn$^{4+}$ ions are not all equal, as assumed in Refs.\
\onlinecite{sessoli,zvezdin}. The Mn$^{4+}$ ions in Mn$_{12}$ 
molecules are arranged in
the four corners of a cubane (distorted cube, see Fig.\ \ref{figcub})
while the other four corners are occupied by
the oxygens providing superexchange bridges between
the manganese ions. Therefore, the manganeses form a distorted tetrahedron
with four short edges of the length 5.33 a.u. and two long edges of the
length of 5.56 a.u., and the angles Mn-O-Mn in the cubane which correspond
to the long and short edges of the manganese tetrahedron are 95.5$^{\circ }$
and 100.6$^{\circ }$. In spite of the moderate difference in the angles and
lengths of the exchange pathways, the calculated coupling parameters turn
out very different; the interactions $J_{3}^{\prime}$ for the longer
pathways are about three times smaller than the interactions $J_3$ for the
shorter ones.

All the previous considerations \cite{zvezdin,Mn12,sessoli} of the exchange
interactions in Mn$_{12}$ were based on the localized Heisenberg model,
assuming that the values of the magnetic moments and the exchange parameters
do not depend on the spin arrangement of the molecule. To check the accuracy
of this assumption, we have performed the calculations for the totally
ferromagnetic spin configuration, when all manganese spins are directed up,
using the parameters $U=8$ eV and $J=0.9$ eV. The results are the following: 
\begin{eqnarray*}
\mu _{\text{Mn1}} &=&3.06\ \mu _{B},\ \mu _{\text{Mn2}}=3.55\ \mu _{B},\ \mu
_{\text{Mn3}}=3.85\ \mu _{B}, \\
\ {\text{gap}} &=&1.63\ {\text{eV}}, \\
J_{1} &=&-36\ {\text{K}},\ J_{2}=-18\ {\text{K}}, \\
J_{3} &=&-22\ {\text{K}},\ J_{3}^{\prime }=+1\ {\text{K}},\ J_{4}=-6\ {\text{%
K}}.
\end{eqnarray*}
Comparing these results with the Tables \ref{momcomp},\ref{exchreal}, we see that
the values of the moments vary by no more than 5\%, but the exchange
integrals $J_1$ and $J_3$ decrease by about 25-30\%, and the exchange $%
J^{\prime}_3$ almost vanishes. These changes are due to the strong covalency
effects and the corresponding partial delocalization of the $3d$-electrons
discussed above.

The energy of the exchange excitations in the Mn$_{12}$ molecule, e.g., the
difference between the ground-state $S=10$ multiplet and the close excited
multiplet, can be crudely obtained from a simple mean-field estimate,
assuming that the pairs Mn$^{4+}$-Mn$^{3+}$ ions form stiff dimers with the
total spin 1/2 (i.e., $J_1\to -\infty$). 
Then, the energy of the ground state
is $E_0 = -2 J_3 - J^{\prime}_3 - 32 J_4 + 16 J_2$ and the energy of the
first excited state is $E_1 = 2 J_3 + J^{\prime}_3$, so that the excitation
energy is 52 K. This number is close to the values 58--66 K obtained in the
neutron scattering experiments \cite{neutron}. Of course, this agreement
should not be taken literally (it is a semiclassical estimate, 
the value of $J_1$ is not big enough, we did
not take into account the contributions of relativistic interactions which
could be comparable to the exchange contribution \cite{Mn12}, etc.), but we
believe that this indicates qualitative adequacy of the calculated exchanges.

\section{Conclusions}

We have applied the LDA+U approach to the study of intramolecular magnetic
interactions in Mn$_{12}$-acetate system. We have shown that the account of
the on-site Coulomb interactions provides a good
description of the main
features of the electronic and magnetic structure, giving the finite energy
gap and the integer value for the magnetic moment of the molecule. The
calculated partial magnetic moments, charge disproportionality between the
Mn1 and Mn2, Mn3 ions, and the energy gap are in agreement with existing
experimental data. The calculations show significant $p$-$d$ hybridization
and considerable covalency effects, and the contribution of delocalized
electrons in the total magnetic moment of the molecule is about 10\%. The
calculated intramolecular exchange interactions are in qualitative agreement
with the existing data. An interesting result of the calculations is that
the interactions between the Mn$^{4+}$ ions situated in the inner cubane are
not all equal, differing by about a factor of three for different
exchange-coupled pairs. The applicability of localized Heisenberg model has
been estimated by comparing the values of the exchange interactions for
different spin configurations (ground-state ferrimagnetic and totally
ferromagnetic), and corresponding variation of the coupling parameters is
about 30\%.

\begin{acknowledgments}
Authors would like to thank A. K. Zvezdin, D. A. Clark, 
and V. V. Platonov for the access
to the unpublished materials on the megagauss-field experiments, and to A.
B. Sushkov and J. L. Musfeldt for providing the unpublished optical data for
Mn$_{12}$. We are grateful to W. Wernsdorfer and P. C. E. Stamp for fruitful
discussions.

The work is partially supported by the Netherlands Organization for
Scientific Research, NWO project 047-008-16. This work was partially carried
out at the Ames Laboratory, which is operated for the U.\ S.\ Department of
Energy by Iowa State University under Contract No.\ W-7405-82 and was
supported by the Director of the Office of Science, Office of Basic Energy
Research of the U.\ S.\ Department of Energy.
This work was partially supported by Russian Basic Research
Foundation, grants 00-15-96544 and 01-02-17063.
\end{acknowledgments}

\begin{table}[b]
\caption{The calculated magnetic moments (in $\protect\mu_B$)
of the
manganese ions, values of the exchange interaction parameters
(in K) and the
energy gap (in eV) for different values of $U$ (in eV). 
The value of $J$ is
0.9 eV. The parameters of the exchange interactions 
$J_3$ correspond to
the exchange between the closest Mn1 ions in the 
cubane (see text).}
\label{momreal}
\begin{ruledtabular}
\begin{tabular}{ccccccccccc}
U & $\mu_{\text{total}}$ & $\mu_{\text{Mn1}}$ & 
$\mu_{\text{Mn2}}$ & $\mu_{%
\text{Mn3}}$ & $J_1$ & $J_2$ & $J_3$ & $J_4$ & gap &  \\ 
\hline 4.0 & 20 & -2.72 & 3.44 & 3.65 & -53 & -47 & -37 &
-19 & 1.35 & 
\\ 
6.0 & 20 & -2.80 & 3.55 & 3.76 & -49 & -37 & -33 & 
-12 & 1.78 &  \\ 
8.0 & 20 & -2.92 & 3.52 & 3.84 & -47 & -26 & 
-30 & -7 & 2.01 & 
\end{tabular}
\end{ruledtabular}
\end{table}

\begin{table}[b]
\caption{The comparison of the calculated magnetic 
moments (in $\protect\mu%
_B $) with the existing theoretical and experimental 
data. {\bf 1} are the
data calculated by Pederson et al., Ref.\ [26,27]; 
{\bf 2} --- the data
calculated by Zeng et al., Ref.\ [28] (only the 
contribution of $d$%
-states is presented); {\bf Exp} --- the experimental results 
of Robinson et
al., Ref.\ [41]. The next two columns are the results of our
calculations done within the LDA and LDA+U methods. LDA+U 
results for the
values $U=8$ eV, $J=0.9$ eV are shown.}
\label{momcomp}
\begin{ruledtabular}
\begin{tabular}{cccccc}
& {\bf 1} & {\bf 2} & {\bf Exp} & LDA & LDA+U \\ 
\hline $\mu_{\text{Mn1}}$ & -2.6 & -3.05 & -2.34 & -2.41 
& -2.92 \\ 
$\mu_{\text{Mn2}}$ & 3.6 & 3.89 & 3.69 & 3.53 & 3.52 \\ 
$\mu_{\text{Mn3}}$ & 3.6 & 4.04 & 3.79 & 3.55 & 3.84 \\ 
$\mu_{\text{total}}$ &  &  & 20.56 & 19.00 & 20.00
\end{tabular}
\end{ruledtabular}
\end{table}

\begin{table}[b]
\caption{The comparison of the calculated values of the 
exchange parameters
(last line of the table) with the existing data. 
{\bf 3} --- the data
suggested by Sessoli et al., Ref.\ [43]; {\bf 4} 
--- the data of
Platonov et al., Ref.\ [23]. LDA+U results for the 
values $U=8$ eV, $%
J=0.9$ eV are shown. The parameters of the exchange 
interactions $J_3$ and $%
J^{\prime}_3$ correspond to the exchange inside different 
pairs of Mn1 ions
in the cubane (see text). All values are given in Kelvin.}
\label{exchreal}
\begin{ruledtabular}
\begin{tabular}{cccccc}
& $J_1$ & $J_2$ & $J_3$ & $J^{\prime}_3$ & $J_4$ \\ 
\hline {\bf 3} & -215 & -86 & -86 & --- & (?)30 \\ 
{\bf 4} & -86 & -20 & -23 & --- & 0.0 \\ 
LDA+U & -47 & -26 & -30 & -10 & -7
\end{tabular}
\end{ruledtabular}
\end{table}


\begin{thebibliography}{}
\bibitem{Molmag}  O. Kahn, {\it Molecular magnetism\/} (VCH, New York, 1993)

\bibitem{Molmagn1}  {\it Molecular magnets: from molecular assemblies to the
devices\/} (Kluwer, Dordrecht, 1996), edited by E. Coronado, P. Delha\`es,
D. Gatteschi, and J. S. Miller.

\bibitem{Molmagn2}  D. Gatteschi, A. Caneschi, L. Pardi, and R. Sessoli,
Science {\bf 265}, 1054 (1994).

\bibitem{QTM}  {\it Quantum tunneling of magnetization - QTM'94},
Vol. 301 of {\it NATO Advanced Study Institute, Series E\/} (Kluwer,
Dordrecht, 1995), edited by L. Gunther and B. Barbara.

\bibitem{QTM1}  E. M. Chudnovsky and L. Gunther, Phys. Rev. Lett. {\bf 60,}
661 (1988)

\bibitem{QTM2}  L. Thomas, F. Lionti, R. Ballou, D. Gatteschi, R. Sessoli,
and B. Barbara, Nature {\bf 383,} 145 (1996)

\bibitem{QTM3}  J. R. Friedman, M. P. Sarachik, J. Tejada, and R. Ziolo,
Phys. Rev. Lett. {\bf 76,} 3830 (1996)

\bibitem{QTM4}  A. Garg, Europhys. Lett. {\bf 50}, 382 (2000)

\bibitem{QTM5}  W. Wernsdorfer, R. Sessoli, A. Caneschi, D. Gatteschi, A.
Cornia, Europhys. Lett. {\bf 50}, 552 (2000).

\bibitem{stamp}  N. V. Prokof'ev and P. C. E. Stamp, Rep. Prog. Phys. {\bf %
63,} 669 (2000).

\bibitem{lis}  T. Lis, Acta Crystallogr. B {\bf 36,} 2042 (1980).

\bibitem{sesnat} R. Sessoli, D. Gatteschi, A. Caneschi, and H. A. Novak,
Nature {\bf 365}, 141 (1993).

\bibitem{rigid}  F. Hartmann-Boutron, P. Politi and J. Villain, Int. J. Mod.
Phys. B {\bf 10}, 2577 (1996).

\bibitem{rigid1}  V. V. Dobrovitski and A. K. Zvezdin, Europhys. Lett. {\bf %
38}, 377 (1997).

\bibitem{rigid2}  L. Gunther, Europhys. Lett. {\bf 39}, 1 (1997).

\bibitem{rigid3}  D. A. Garanin and E. M. Chudnovsky, Phys. Rev. {\bf B 56},
11 102 (1997).

\bibitem{rigid4}  A. Fort, A. Rettori, J. Villain, D. Gatteschi and R.
Sessoli, Phys. Rev. Lett. {\bf 80}, 612 (1998).

\bibitem{rigid5}  F. Luis, J. Bartolome, and J. F. Fernandez, Phys. Rev. B 
{\bf 57}, 505 (1998).

\bibitem{rigid6}  N. V. Prokof'ev and P. C. E. Stamp, Phys. Rev. Lett. {\bf %
80}, 5794 (1998).

\bibitem{rigid7}  K. Saito, S. Miyashita, and H. De Raedt, Phys. Rev. B {\bf %
60}, 14553 (1999).

\bibitem{rigid8}  A. Cuccoli, A. Fort, A. Rettori, E. Adam, and J. Villain,
Europ. Phys. Journ. B {\bf 12}, 39 (1999).

\bibitem{rigid9}  M. N. Leuenberger and D. Loss, Phys. Rev. B {\bf 61}, 1286
(2000).

\bibitem{neutron}  M. Hennion, L. Pardi, I. Mirebeau, E. Suard, R. Sessoli,
and A. Caneschi, Phys. Rev. B {\bf 56,} 8819 (1997); I. Mirebeau, M.
Hennion, H. Casalta, H. Andres, H. U. Gudel, A. V. Irodova, and A. Caneschi,
Phys. Rev. Lett. {\bf 83,} 628 (1999).

\bibitem{zvezdin}  V. V. Platonov, B. Barbara, A. Caneschi, D. A. Clark, C.
M. Fowler, D. Gatteschi, J. D. Goettee, I. A. Lubashevsky, A. A. Mukhin, V.
I. Plis, A. I. Popov, D. G. Rickel, R. Sessoli, O. M. Tatsenko, and A. K.
Zvezdin, (to be published); B. Barbara, D. Gatteschi, A. A. Mukhin,
V. V. Platonov, A. I. Popov, A. M. Tatsenko, A. K. Zvezdin, 
Proceeding of Seventh International Conference on Megagauss Magnetic
Field Generation and Related Topics (Sarov, 1996), 853 (1997).

\bibitem{optics}  S. M. Oppenheimer, A. B. Sushkov, J. L. Musfeldt, R. M.
Achey, and N. S. Dalal, (submitted to Phys. Rev. B).

\bibitem{Mn12}  M. I. Katsnelson, V. V. Dobrovitski, and B. N. Harmon, Phys.
Rev. B {\bf 59,} 6919 (1999).

\bibitem{ellis}  Z. Zeng, D. Guenzburger, and D. E. Ellis, Phys. Rev. B {\bf %
59,} 6927 (1999).

\bibitem{Ped1}  M. R. Pederson and S. N. Khanna, Chem. Phys. Lett. {\bf 307,}
253 (1999); Phys. Rev. B {\bf 59,} 9566 (1999).

\bibitem{Ped2}  M. R. Pederson and S. N. Khanna, Phys. Rev. B {\bf 60,} 9566
(1999).

\bibitem{v15exch}  J. Kortus, C. S. Hellberg, and M. R. Pederson, Phys. Rev.
Lett. {\bf 86,} 3400 (2000).

\bibitem{oxide}  N. F. Mott, {\it Metal-Insulator Transitions} (Taylor and
Francis, London, 1974)

\bibitem{oxide1}  K. Terakura, A. R. Williams, T. Oguchi, J. K\"ubler, Phys.
Rev. Lett. {\bf 52}, 1830 (1984)

\bibitem{oxide2}  A. Svane and O. Gunnarsson, Phys. Rev. Lett. {\bf 65},
1148 (1990)

\bibitem{oxide3}  A. I. Lichtenstein and M. I. Katsnelson, Phys. Rev. B {\bf %
57}, 6884 (1998)

\bibitem{oxide4}  K. Held, G. Keller, V. Eyert, D. Vollhardt, and V. I.
Anisimov, Phys. Rev. Lett. {\bf 86,} 5345 (2001).

\bibitem{LDA+U}  V. I. Anisimov, F. Aryasetiawan, and A. I. Lichtenstein, J.
Phys.: Condens. Matter {\bf 9,} 767 (1997).

\bibitem{MnO}  I. V. Solovyev and K. Terakura, Phys. Rev. B {\bf 58,} 15496
(1998).

\bibitem{MnO1}  I.A. Nekrasov, M.A. Korotin, and V.I. Anisimov,
cond-mat/0009107.

\bibitem{lmto1}  O. Gunnarsson, O. Jepsen, and O. K. Andersen, Phys. Rev. B 
{\bf 27}, 7144 (1983).

\bibitem{KL}  A. I. Liechtenstein, M. I. Katsnelson, V. P. Antropov, and V.
A. Gubanov, J. Magn. Magn. Mater. {\bf 67,} 65 (1987); M. I. Katsnelson and
A. I. Lichtenstein, Phys. Rev. B {\bf 61,} 2230 (2000).

\bibitem{LAZ}  A. I. Liechtenstein, V. I. Anisimov, and J. Zaanen, Phys.
Rev. B {\bf 52,} R5467 (1995).

\bibitem{rob}  R. A. Robinson, P. J. Brown, D. N. Argyriou, D. N.
Hendrickson, S. M. J. Aubin, J. Phys.: Condens. Mat. {\bf 12}, 2805 (2000).

\bibitem{mnxps}  P. Ghigna, A. Campana, A. Lascialfari, A. Caneschi, D.
Gatteschi, A. Tagliaferri, and N. Brookes, cond-mat/0007092.

\bibitem{sessoli}  R. Sessoli, H.-L. Tsai, A. R. Shake, S. Wang, J. B.
Vincent, K. Folting, D. Gatteschi, G. Christou, D. N. Hendrickson, J. Am.
Chem. Soc. \textbf{115}, 1804 (1993).

\bibitem{cheesman}  M. R. Cheesman, V. S. Oganesyan, R. Sessoli, D.
Gatteschi, and A. J. Thomson, Chem. Commun. {\bf 17}, 1677 (1997).
\end{thebibliography}
\end{document}